# Presence of pKa Perturbations Among Homeodomain Residues Facilitates DNA Binding

**Christopher M. Frenz[1*] and Philippe P. Lefebvre[1]**
[1]Department of Otolaryngology
University of Liege
CHU de Liege
4000 Liege, Belgium
[*]Corresponding Author: cfrenz@gmail.com

**Abstract** - *Homeodomain containing proteins are a broad class of DNA binding proteins that are believed to primarily function as transcription factors. Electrostatics interactions have been demonstrated to be critical for the binding of the homeodomain to DNA. An examination of the electrostatic state of homeodomain residues involved in DNA phosphate binding has demonstrated the conserved presence of upward shifted pKa values among the basic residue of lysine and arginine. It is believed that these pKa perturbation work to facilitate binding to DNA since they ensure that the basic residues always retain a positive charge*.

## 1 Introduction

Homeobox proteins are characterized by a 60 amino acid helix-turn-helix motif that is involved in the binding of DNA. Homeobox proteins have been characterized as transcription factors that regulate gene expression during the development and differentiation of neural systems within mammals, although they are also present in a wide range of eukaryotes and in bacteria as repressor proteins [1,2]. Recent examinations of the forces involved in the binding of the homeodomains of Antp, NK2, Engrailed, and Matα2 to DNA have demonstrated that DNA binding is largely an electrostatic phenomena. Unbound homeodomains were demonstrated to have chloride ions bound to the residues that will coordinate with the phosphates of DNA upon binding. During the binding process these chloride ions are displaced along with calcium ions that were bound to the DNA phosphates, and the ion interactions replaced by electrostatic coordination between homeodomain residues and the DNA phosphates [2].

Electrostatics playing a key role in ligand binding is not limited to homeodomain containing proteins, however, and electrostatic perturbations, as demonstrated by a pKa shift of greater than 2 pK units, have been recently demonstrated among residues involved in ligand binding. Within protein-protein complexes 80% of acidic residues have been demonstrated to have a negatively shifted pKa value that allows residues to maintain their negatively charged state within the complex [3]. Within the HCV NS3 helicase, the glutamate residue at position 493 was demonstrated to have a positive shifted perturbation which enabled the glutamate to retain a neutral charge and to coordinate with DNA. Perturbations were also demonstrated within the site of ATP binding and hydolysis [4] for the HCV helicase, a DEXH motif containing region of the protein, while another study has demonstrated that the related DEAD-box motif exhibited conserved electrostatic properties among multiple DEAD-box proteins [5].

Moreover, recent work has suggested that protein electrostatic states are evolutionary conserved. Conserved electrostatic interaction networks have been demonstrated in TIM-barrel proteins and the electrostatic surfaces of proteins from 4 distinct protein families and 1 superfamily have been demonstrated to be evolutionary conserved as well [6,7]. Studies of the HIV reverse transcriptase and the HCV helicase and polymerase have further demonstrated that the greatest number and magnitude of electrostatic perturbations exist in the residues that are most highly conserved [8]. Additionally, an electrostatic investigation of nucleoside monophosphate kinase proteins determined that the pKa values of highly conserved titratable groups within the protein families tended to also have highly conserved pKa values [9]. Moreover, the electrostatic potential maps of lysozyme proteins from a variety of organism sources were demonstrated to have a high degree of similarity, suggesting the evolutionary conservation of the electrostatic state of the protein [10].

Given the importance of electrostatics to the DNA binding of homeodomain containing proteins, this study seeks to determine if any electrostatic perturbations are present among the DNA phosphate binding residues of homeobox proteins and to determine whether pKa perturbations are conserved among a diversity of homeobox proteins.

## 2 Methods

### 2.1 Sequence and Structure Selection

Proteins for inclusion in this study were selected by performing a BLAST search for the homeodomain of the Oct-

1 transcription factor against the PDB sequence database [11] to ensure that only proteins with crystal structures, a requirement for the electrostatic calculations, were identified. Trying the BLAST search with other homeodomain containing sequences yielded similar BLAST results. Sixty three protein sequences were returned and sequence alignment was performed with MAFFT [12]. Of the 63 proteins available, 15 proteins bound to DNA were chosen for electrostatic calculations and pKa determination (Table 1). The few DNA bound proteins not selected were due to missing residues being present in the structures, which the H++ server lacks the capability of processing.

**Table 1:** Homeodomain protein structures on which electrostatic computations were performed in order to determine protein pKa values.

| PDB ID | Protein | Source |
|---|---|---|
| **1AHD** | ANTENNAPEDIA HOMEODOMAIN-DNA COMPLEX | Billeter et al., 1993 |
| **1AKH** | MAT A1/ALPHA2/DNA TERNARY COMPLEX | Li et al., 1998 |
| **1OCT** | OCT-1 | Klemm et al., 1994 |
| **1YRN** | MATA1/MATALPHA2 HOMEODOMAIN | Li et al., 1995 |
| **1ZQ3** | BICOID HOMEODOMAIN | Baird-Titus et al., 2006 |
| **2HOs** | PHAGE-SELECTED HOMEODOMAIN | Shokat et al., 2006 |
| **9ANT** | ANTENNAPEDIA HOMEODOMAIN-DNA COMPLEX | Fraenkel & Pabo, 1998 |
| **1FJL** | DROSOPHILA PAIRED PROTEIN | Wilson et al., 1995 |
| **1NK3** | VND/NK-2 HOMEODOMAIN/DNA COMPLEX | Gruschus et al., 1997 |
| **1DU0** | ENGRAILED HOMEODOMAIN | Grant et al., 2000 |
| **1LE8** | MATa1/MATalpha2-3A HETERODIMER | Ke et al., 2002 |
| **1HDD** | ENGRAILED HOMEODOMAIN | Kissinger et al., 1990 |
| **2H1K** | Pdx1 HOMEODOMAIN | Longo et al., 2007 |
| **1PUF** | HoxA9 HOMEODOMAIN | Laronde-Leblanc & Wolberger, 2003 |
| **1LFU** | PBX HOMEODOMAIN | Sprules et al., 2003 |

Based on some preliminary findings, non-DNA bound homeodomains were not considered accurate electrostatic representations since chloride ion coordination has been demonstrated to be crucial to the unbound forms of the proteins. The explicit placement of these ions is not available in the structural data and the ions are thus not feasible to accurately account for their effects in the pKa calculations. The importance of negative charges to the electrostatics of homeodomains is further supported by calculations where the bound DNA ligands were removed, which resulted in a loss of perturbation among the DNA binding residues.

### 2.2 Electrostatic Calculations

pKa values were calculated using the H++ Web Server, available at http://biophysics.cs.vt.edu/H++ [13]. pKa predictions are begun by adding missing atoms and assigning partial charges to the uploaded protein structure using the parm99 force field and the AMBER molecular modeling package [14]. Positions of these added protons are optimized using 100 steps of conjugate gradient descent minimization and 500 steps of Molecular Dynamics simulation at 300K. The Poisson-Boltzmann equation in the program package MEAD is used to compute the free energies of the protonation microstates [15]. Titration curves and pKa values are then determined using the clustering approach described by Gilson [16]. Perturbations present on the Lysine and Arginine residues involved in DNA phosphate binding were computed by subtracting the standard pKa value of an unperturbed form of the residue (K=10.52, R=12.48) from the calculated value.

### 2.3 Plots and Statistics

Plots and statistics were computed using the program Graph Pad Prism version 4.01.

## 3 Results and Discussion

An amino acid sequence alignment for the homeodomain containing protein sequences was performed and the sequence region identified by Dragan et al. [2] as participating in coordinating with the DNA phosphates located within the alignment identified (Figure 1), as indicated by the residues aligned with the FCNRRQKEKR in the top sequence. This region contains several positions for which the presence of basic amino acid residues (K and R) is conserved. It is these basic residues which have been demonstrated to have interaction with the DNA phosphate groups.

**Figure 1:** BLAST alignment of homeodomain DNA binding sequences.

```
gi|1127240|pdb|  ---------------------------------------FCNRRQKEKR-----------
gi|999603|pdb|1  ---------------------------------------FCNRRQKEKR-----------
gi|1127090|pdb|  ---------------------------------------FCNRRQKGKR-----------
gi|24159066|pdb  ---------------------------------------FINKRMRSK------------
gi|1310757|pdb|  ---------------------------------------FINKRMRSK------------
gi|21466066|pdb  ---------------------------------------FINKRMRSK------------
gi|3212445|pdb|  ---------------------------------------FINKRMRSK------------
gi|34810086|pdb  ---------------------------------------FQNRRMKMKK-----------
gi|4929895|pdb|  ---------------------------------------FQNRRMKLKK-----------
gi|1421276|pdb|  ---------------------------------------FQNRRMKWKK-----------
gi|1431670|pdb|  ---------------------------------------FQNRRMKWKK-----------
gi|443020|pdb|1  ---------------------------------------FQNRRMKWKK-----------
gi|3891759|pdb|  ---------------------------------------FQNRRMKWKK-----------
gi|1421399|pdb|  ---------------------------------------FQNRRMKSKK-----------
gi|4558069|pdb|  ---------------------------------------FQNRRMKQKK-----------
gi|14278377|pdb  ---------------------------------------FQNRRAKAKR-----------
gi|118137506|pd  ---------------------------------------FQNKRSKIKK-----------
gi|118137582|pd  ---------------------------------------YQNRRMKWKK-----------
gi|119389348|pd  ---------------------------------------FQNRRAKWRR-----------
gi|9954899|pdb|  ---------------------------------------FANKRAKIKK-----------
gi|3212684|pdb|  ---------------------------------------FKNKRAKIKK-----------
gi|229970|pdb|1  ---------------------------------------FQNKRAKIKK-----------
gi|38492759|pdb  ---------------------------------------FQNERAKIKK-----------
gi|38492755|pdb  ---------------------------------------FQNARAKIKK-----------
gi|82407634|pdb  ---------------------------------------FQNKRAKIRR-----------
gi|3891326|pdb|  ---------------------------------------FQNKRAKIKK-----------
gi|640374|pdb|1  ---------------------------------------FQNKRAKI-------------
gi|122920443|pd  ---------------------------------------FKNMRAKIKK-----------
gi|83754790|pdb  ---------------------------------------FQNRRVKEKK-----------
gi|118137583|pd  ---------------------------------------FKNRRAKWRR-----------
gi|83754871|pdb  ---------------------------------------FSNRRAKWRREEKLRNQ---R
gi|118137581|pd  ---------------------------------------FKNRRAKCRQQQQQQQN---G
gi|15826439|pdb  ---------------------------------------FQNRRMKDKR-----------
gi|1943465|pdb|  ---------------------------------------FQNHRYKTKR-----------
gi|5542465|pdb|  ---------------------------------------FQNHRYKTKR-----------
gi|1310870|pdb|  ---------------------------------------FQNHRYKMKR-----------
gi|1633405|pdb|  ---------------------------------------FQNRRARLRK-----------
gi|66361272|pdb  ---------------------------------------FKNRRAKWRK-----------
gi|110591143|pd  ---------------------------------------FQNTRARERK-----------
gi|110591142|pd  ---------------------------------------FQNARQKARK-----------
gi|110591141|pd  ---------------------------------------FRNTLFKERQ-----------
gi|90108646|pdb  ---------------------------------------FKNRRRRHKI-----------
gi|110591145|pd  ---------------------------------------FSERRKKVNAEE--------T
gi|118137579|pd  ---------------------------------------FSERRKLRDSMEQAVLDSMGS
gi|118137589|pd  ---------------------------------------FSDRRYHCRNLK--------G
gi|122919811|pd  ---------------------------------------FEQRKVYQYS----------N
gi|5542140|pdb|  ---------------------------------------FQNKRCKDKK-----------
gi|83754779|pdb  ---------------------------------------FRHRRNQDKP-----------
gi|82407413|pdb  ---------------------------------------FRQRRNQEKP-----------
gi|110591144|pd  ---------------------------------------IGNRRRKYRL--------MGI
gi|119389124|pd  ---------------------------------------FMNARRRSLDKWLEHHHHHH-
gi|58176704|pdb  ---------------------------------------FMNAR----------------
gi|119389336|pd  ---------------------------------------LPDELRVEKG-----------
gi|52696088|pdb  LDLAADKIARGIYSYAGQRCDAIKLVLAERPVYGKLVEEVAKRLSSLRVGDPRDPTVDVG
gi|52696082|pdb  LDLAADKIARGIYSYAGQRCDAIKLVLAERPVYGKLVEEVAKRLSSLRVGDPRDPTVDVG
gi|28373511|pdb  LDLAADKIARGIYSYAGQRCDAIKLVLAERPVYGKLVEEVAKRLSSLRVGDPRDPTVDVG
                                                   .
```

**Figure 2 :** pKa perturbations associated with the DNA binding sequences of homeodomain proteins.

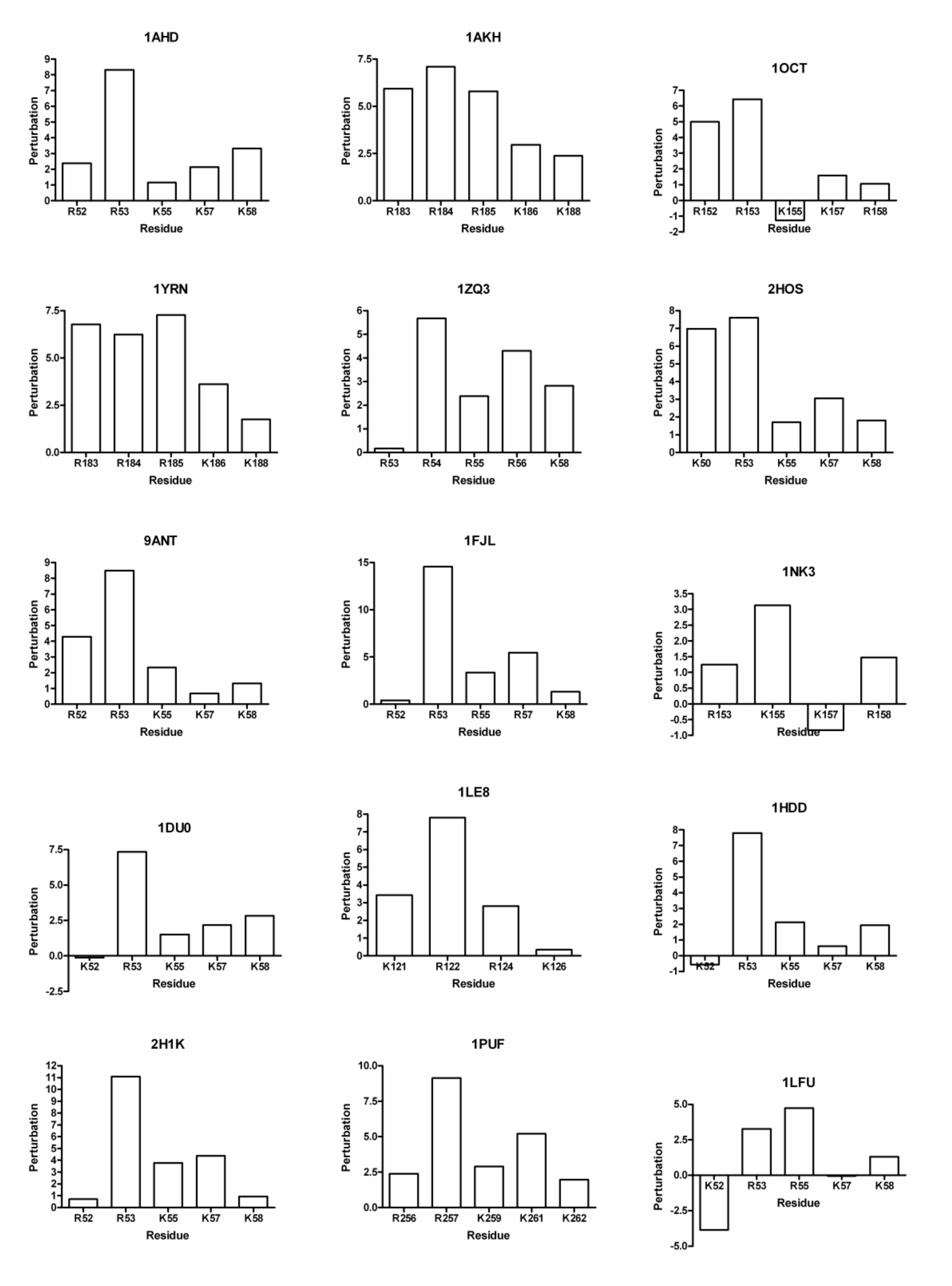

Calculation of the pKa values of the various homeodomain proteins revealed several notable similarities (Figure 2). One such similarity is that in all of the homeobox proteins examined, upward shifted pKa perturbations of at least 2 pK units are presented in at least one of the lysine or arginine residues involved in coordination with the DNA phosphates. Given the high degree of residue conservation among this sequence region of homeobox proteins and the uniformity of the data computed it is hypothesized that these upwards shifted pKa perturbations are also likely conserved among all homeodomain proteins. This hypothesis is further supported by findings that indicate that pKa perturbations tend to only be present amongst residues that are highly conserved [8]. Moreover, the presence of these perturbations makes functional sense, since the upwards shift in pKa, some to pH levels that could never be reached in a physiological system, work to ensure that the lysine and arginine residues always remain in a protonated state, and hence always retain a positive charge. Given the negative charge of DNA the ability to maintain a strong positive charge would work to enhance the ability of these residues to bind DNA. Charge distributions such as this have been demonstrated to be conserved within protein families and for the CuZnSOD family it was suggested that a conserved charge distribution present around the catalytic sites of the enzymes served as a cationic funnel that helped steer the ligand into the catalytic binding site [5]. These findings further support the idea that a conserved electrostatic perturbation may exist in homeodomain structures to facilitate the binding of DNA.

What is also notable is that in all structures analyzed a significant (>2 pK units) perturbation was consistently present in the basic residue that aligned with the second arginine of the FCNRRQKEKR base sequence. In many cases this perturbation was the strongest perturbation present among the DNA phosphate binding residues. One explanation for this is that for the 1NK3 and 9ANT structures it has been reported that out of all of the DNA phosphate binding residues, these residues were the ones that were buried most deeply into the bound DNA [2]. Given that pKa perturbations have been demonstrated to have an association to residue packing, where a tighter packing density has a higher magnitude of perturbation [17], the extensive burying of this residue can help to account for the magnitude of perturbation at this residue position. This suggestion is further supported by the finding that if the DNA is removed from the pKa calculations the observed perturbations diminish back into the range of values typically associated with basic residues. It is further hypothesized that the chloride ions which have been demonstrated to coordinate with the basic residues in the absence of DNA [2] help to maintain this perturbed state in the unbound structures as well, in order to facilitate the binding of DNA even further.

While there appears to be a conservation of the occurrence of an upwardly shifted pKa value for at least one basic residue in each structure it is notable that the positions and the magnitudes of each perturbation are not perfectly conserved. These differences are believed to be the result of the differences in the homeodomain sequences and along with geometrical considerations are believed to contribute to the sequence specificity associated with each type of homeodomain.